# Constraint Clustering Based Islanding Scheme For Power Networks

Hamzeh Davarikia
Electrical and Computer Engineering Department
McNeese State University
Lake Charles, LA, USA
hdavarikia@mcneese.edu

Faycal Znidi
Electrical and Computer Engineering Department
Texas A&M University
Texarkana, TX, USA
fznidi@tamut.edu

Masoud Barati
Electrical and Computer Engineering Department
University of Pittsburgh
Pittsburgh, PA, USA
masoud.barati@pitt.edu

***Abstract*—Controlled islanding, which splits the whole power system into islands, is an effective strategy against rolling blackout during severe disturbances. Finding the islanding solutions in a real-time manner is complicated because of the combinatorial explosion of the solution space occurs for a large power network. In this work, a computationally efficient controlled islanding algorithm is proposed that uses constrained spectral clustering while addressing the generator coherency problem. The objective function used in this controlled islanding algorithm is the minimal power-flow disruption. The sole constraint applied to this solution is related to generator coherency. An undirected edge-weighted graph is created based on absolute values of apparent power flow and constraints related to transmission line availability and coherent generator groups are included by altering the edge weights of the graph and using a subspace projection. Spectral clustering is then applied to the constrained solution subspace to find the islanding solution. The methodology is tested on an IEEE-39 test system with a fully dynamic model. Simulation results demonstrate the efficacy of our approach.**

## I. Introduction

Power systems subjected to large disturbances may become transiently unstable leading to uncontrolled system separation due to cascading failures. Controlled islanding can help to prevent the detrimental impacts of large disturbances and contain the impact of the disturbance to a smaller island where the service would only be slightly degraded. Most approaches to islanding aim to identify as primary objective stable islands with the minimal load shedding. To find reasonable islanding cutsets, all subsystems must satisfy some set of attributes such as generator coherency, power flow disruption, voltage limits, transmission line thermal limits, transient stability, etc. [1]. This set of attributes can be coordinated with other corrective measures to find applicable islands in the power system that will help to reduce the complexity of the islanding solution, which is similar to the 0-1 knapsack problem [2].

Generally, selecting the best option among a set of alternatives is not always a straightforward process. The typical methodology to solve this problem is to consider every existing combination of attributes and then attempt to rank all the combinations. However, the problem rapidly becomes intractable, as the number of possible combinations grows exponentially as the number of attributes increases in the so-called NP-hard problems. Instead, we may choose the alternative that contains all those highest ranked attributes, which have the greatest impact on the islanding boundaries' formation. This set of attributes can be coordinated with other corrective measures to find applicable islands in the power system that will help to reduce the complexity of the *ICI*, which is similar to the 0-1 knapsack problem [2].

Traditionally, the islanding problem has been solved using combinatorial optimization approaches. In [1], a Binary Particle Swarm Optimization (BPSO) seeking Pareto Non-dominated solutions algorithm is presented to find islands with minimal power imbalances that contain coherent generators. In [2], the Mixed Integer Linear Programming (MILP) algorithm is used to find the islanding solution that detaches coherent generator groups with a minimal power imbalance. In [3], the authors present a technique for system splitting by utilizing the OBDD method. In the situation of the splitting system into two islands, each load bus either belongs to one island or the other. In [4], a mixed-integer linear programming problem based splitting strategies is presented to handle energy production and demand. In [5], an approach based on dynamic frequency deviations of both generator and non-generator buses, concerning the system nominal frequency is presented. Overall the center of the inertia concept has shown its advantages in various applications.

Several approached used the spectral clustering to find an islanding solution. In [6, 7], a constrained spectral k-embedded clustering method is defined to find an islanding scheme with minimal active power flow disruption while addressing the generator's coherency problem. In [8], a two-step Spectral Clustering Controlled Islanding algorithm (SCCI) is presented while using generator coherency as the sole constraint with minimum active power flow disruption to find a suitable ICI solution. These techniques disregard the effects of the bus voltage magnitude, and reactive power, which has a substantial impact on the dynamic coupling. In these methodologies the reactive power is viewed as a local issue and can be handled with local reactive power compensators and only active power is considered in the splitting scheme. However, the reactive power plays a significant role in supporting the voltage profile, and a significant mismatch of the reactive power supply and demand causes high or low voltage conditions within islands [9]. Another interesting cognitive approach that can be used toward solving the problem of groups of generators can be obtained from [10,11], where the author analyses the behavior of an individual in a group in a collaborative environment.

In this work, a controlled islanding method based on constrained spectral clustering while addressing the generator coherency problem is presented. Further, the proposed methodology was designed and implemented to offer high-speed performance without the need to reduce the size of the power system model. The objective function used in this controlled islanding algorithm is the minimal power-flow disruption. The sole constraint applied to this solution is related to generator coherency. An undirected edge-weighted graph is created based on absolute values of apparent power flow and constraints related to transmission line availability and coherent generator groups are included by altering the edge weights of the graph and using a subspace projection.

## II. Dynamic Coherency Detection

### A. Synchronization power coefficient as coherency index between generators

The synchronous generators are the most significant players in the power system dynamics. The synchronization phenomenon of generator groups offers invaluable information about system behavior. The coherency among the generators, i.e., their tendency to swing together, changes following a disturbance, a characteristic that is completely considered in this methodology in solving the islanding problem to ensure the wide-area stability of the islanded system by creating self-operating and stable island grids. Znidi et.al [1] proposed an algebraic model for calculating the synchronizing power coefficient for a multi-machine system between $m$ generators in $n$ bus system with classical machine models, constant impedance loads, and the network reduced to generator internal nodes. Equation (1) shows the synchronization power coefficient between generator $i$ and $j$.

$$\frac{\partial P_{ij}}{\partial \delta_{ij}} = P_{syc_{ij}} = \sum_{j \in \mathcal{M}}^{m} |E_i'||E_j'|(-B_{ij}\cos\delta_{ij}) \tag{1}$$

where $|E_i'|$ is the magnitude of the voltage behind the sub-transient reactance in the synchronous generator, $B_{ij}$ is the susceptance between the element $i$ and $j$ in the reduced system, and $\delta_{ij}$ is the difference between the rotor angle of machines $i$ and $j$ [12-14]. A small change in powers at generator $i$ and $j$, causing a small change in angular separation between generators $i$ and $j$. Using the undirected graph, there is an associated complete weighted graph such that:

$$P_{syc_m} = (G\ , P_{syc}) \ni P_{syc} = \left\{ P_{syc_{ij}} |\ i,j \in\ G\ , i \neq j \right\} \tag{2}$$

where $m$ is the number of machines in the network and $G = \{G_1, \dots, G_\mathrm{m}\}$ is the set of generators and $P_{syc}$ is the set of synchronization power coefficient among the generators. The associated adjacency matrix to the $P_{syc_m}$ is is a square $m \times m$ matrix that can be easily formed in real-time fashion for a power network.

### B. Coherency Identification based on Spectral Clustering

Given the matrix of the synchronization power coefficient, (2), the identification of the coherent groups for $m$ generators is regarded as an *N-cut problem.* Spectral clustering has become a prevalent clustering method in the past few years [14]. In graph theory, spectral clustering is equivalent to minimizing weights of graph cuts, which can be solved by the normalized cuts algorithm. Therefore, spectral clustering groups the generators into coherent groups based on the eigenvector associated with the smallest eigenvalue of the normalized Laplacian of the graph. The objective in this method is to separate the network into groups of vertices that have weak intergroup connections and strong intragroup connections, by looking for naturally occurring groups in a network regardless of their number or size. It essentially corresponds to the eigenvalues and eigenvectors of the normalized graph Laplacian. The Laplacian is a symmetrical matrix defined as:

$$L_i = D^{\frac{1}{2}}(D - W)D^{-\frac{1}{2}} \tag{3}$$

where $D$ is the degree matrix containing a degree of vertices along diagonal and its element $D_{ij} = \sum_{j=1}^{M} P_{syc_m}$ and $W$ is the adjacency matrix and $P_{syc_{ij}}$ is the element of the adjacency matrix $P_{syc_m}$ of the power graph.

Therefore, by using (1) and constructing (2), it is possible to identify the coherent groups based on an equivalent graph of the synchronization power coefficient among generators using the spectral clustering algorithm.

## III. Controlled Islanding and Constrained Spectral Clustering

In this methodology, the controlled islanding problem is defined as a constrained combinatorial optimization problem and a graph-cut problem. A feasible technique for solving this optimization problem, constrained spectral clustering, is then introduced.

To form a stable island several key attributes must be considered, i.e., generator coherency, load-generation balance, and transient stability [10]. The controlled islanding problem is, in mathematical terms, a multi-objective, multi-constraint non-linear optimization problem which is generally very difficult to solve. The inclusion of reactive power or voltage in the constraints results in a mixed-integer nonlinear program (MINLP), in the so-called NP-hard problems [15]. Consequently, several methodologies ignore the effects of the reactive power and bus voltage magnitude, which has an impact on the dynamic coupling [16].

Mostly, in a power system, the voltage and the frequency are controlled by reactive power (Q) and active power (P), respectively. Consequently, by considering P and Q simultaneously in the islanding problem, the result would be more stable islands in terms of frequency and voltage. Active and reactive power balance plays an important role in stable splitting strategies for island operation of power systems and maintains an acceptable voltage and frequency profile in the post-islanding period [17-20]. The active power and reactive power together make up apparent power as shown in (4).

$$S_{ij} = \sqrt{{P_{ij}}^2 + {Q_{ij}}^2} \tag{4}$$

where $P_{ij}$, $Q_{ij}$, and $S_{ij}$ are the absolute active, reactive, and apparent power flow, respectively between bus $i$ and $j$

The power network can be represented based on the graph-model **G**(**V**, **E**, **W**) of an *n-bus* power network. In this model, the edge set **E**, with elements $e_{ij} = (i, j = 1, \ldots, n)$, represent the transmission lines and the node-set $\boldsymbol{V} = \{v_1, \ldots, v\}$ represents the buses and **W** is a set of edge weights.

The controlled islanding problem can be characterized as constrained combinatorial optimization, with the objective function provided in (5). This objective function minimizes the sum of the absolute values of the apparent power flow between islands, which is the *power-flow disruption*. The constraints applied when satisfying this objective function deal with coherent generator groups and transmission line availability.

$$\min_{\boldsymbol{V}_1,\, \boldsymbol{V}_2 \subset \boldsymbol{V}} \left( \sum_{i \in \boldsymbol{V}_1, j \in \boldsymbol{V}_2} |S_{ij}| \right) \quad St. \tag{5}$$
$$\boldsymbol{V}_{G1} \subset \mathbf{V}_1,\ \boldsymbol{V}_{G2} \subset \mathbf{V}_2, \boldsymbol{E}_s \neq 0 \ and \ \boldsymbol{E}_s \subset \boldsymbol{E}$$

where $S_{ij}$ represents the absolute value of the apparent power flow between node *i* and *j* and $\boldsymbol{E}_s$ is an edge subset that contains all the transmission lines that cannot be separated.

The controlled islanding problem with the above objective functions can be transformed into a *graph-cut problem* by defining a squared $n \times n$ adjacency matrix whose elements, $S_{ij}$ are the absolute apparent power flow between bus *i* and *j*. The graph **G** can be divided into sub-graphs $\boldsymbol{G}_1(\boldsymbol{V}_1, \boldsymbol{E}_1, \boldsymbol{W}_1)$ and $\boldsymbol{G}_2(\boldsymbol{V}_2, \boldsymbol{E}_2, \boldsymbol{W}_2)$. Where $\boldsymbol{V}_1$ and $\boldsymbol{V}_2$ are two disjoint sub-sets representing the buses in island 1 and 2, while $\boldsymbol{V}_{G1}$ and $\boldsymbol{V}_{G2}$ and are the corresponding subsets of coherent generators within each island. To solve the graph-cut problem, graph **G** must be partitioning into two sub-graphs $\boldsymbol{G}_1$ and $\boldsymbol{G}_2$ by removing the edges connecting $\boldsymbol{G}_1$ and $\boldsymbol{G}_2$. The sum of the weights of the cutset is called cut, which is described as:

$$Cut(\boldsymbol{V}_1, \boldsymbol{V}_2) = \sum_{i \in \boldsymbol{V}_1, j \in \boldsymbol{V}_2} w_{ij} \tag{6}$$

Therefore, the controlled islanding problem (5) is transformed into the problem of finding the cutset that divides a graph with a minimum cut. The controlled islanding problem can be assessed as a graph-cut problem utilizing constrained spectral clustering. In the constrained spectral clustering two types of constraints can be defined as Must Link (ML) and Cannot Link (CL)

1. ML-constraint between two vertices assures that those vertices will be in the same cluster.
2. CL-constraint guarantees that the vertices will be in different clusters.

To apply constraints into spectral clustering problem, constraint projection matrix $Q$ is defined as below:

$$Q = \begin{cases} 1 & if(x_i, x_j) \in CL \\ -1 & if(x_i, x_j) \in ML \\ 0 & else \end{cases} \tag{7}$$

With the introduction of the projection matrix, constrained spectral clustering can be applied to the undirected graph to find the cutset with the minimal power-flow disruption that satisfies the generator grouping constraints produced in the first step. Below shows a step-by-step list of instructions to implement the proposed methodology:

1. Obtain k (total number of clusters) based on generator coherency information.
2. Construct the edge weight matrix using the normalized Laplacian matrix $L_i$ introduced in (3).
3. Construct the projection matrix **Q** using generator coherency information.
4. Compute the first two eigenvectors $u_1$ and $u_2$, of the generalized Eigen problem $Q^T \boldsymbol{L}_i \boldsymbol{Q} u = \lambda \boldsymbol{Q}^T \boldsymbol{Q} u$.
5. Let $\boldsymbol{J} \in \boldsymbol{\mathcal{R}}^{n \times 2}$ be the matrix containing the vectors $Qu_1$, $Qu_2$ as columns let $y_i \in \boldsymbol{\mathcal{R}}^{n \times 2}$ be the vector corresponding to the i[th] row of $\boldsymbol{J}$.
6. Cluster the nodes $y_i \in \boldsymbol{\mathcal{R}}^2$ into the clusters $\boldsymbol{V}_1$, $\boldsymbol{V}_2$ using the k-medoids algorithm.

## IV. Simulation Test Cases

The model effectiveness is evaluated through the simulation study conducted on the modified IEEE 39-bus system. The methodology has been implemented in MATLAB and all time-domain simulations are achieved in DIgSILENT PowerFactory. Table 1 list the events that occurred as a result.

Table 1. Events Occurred in Case I

| Time (sec) | Description |
|---|---|
| 2.00 | Short circuit on lines 3-4 |
| 2.40 | Switch event on lines 3-4 |
| 3.00 | Short circuit on lines 16-17 |
| 3.40 | Switch event on lines 16-17 |

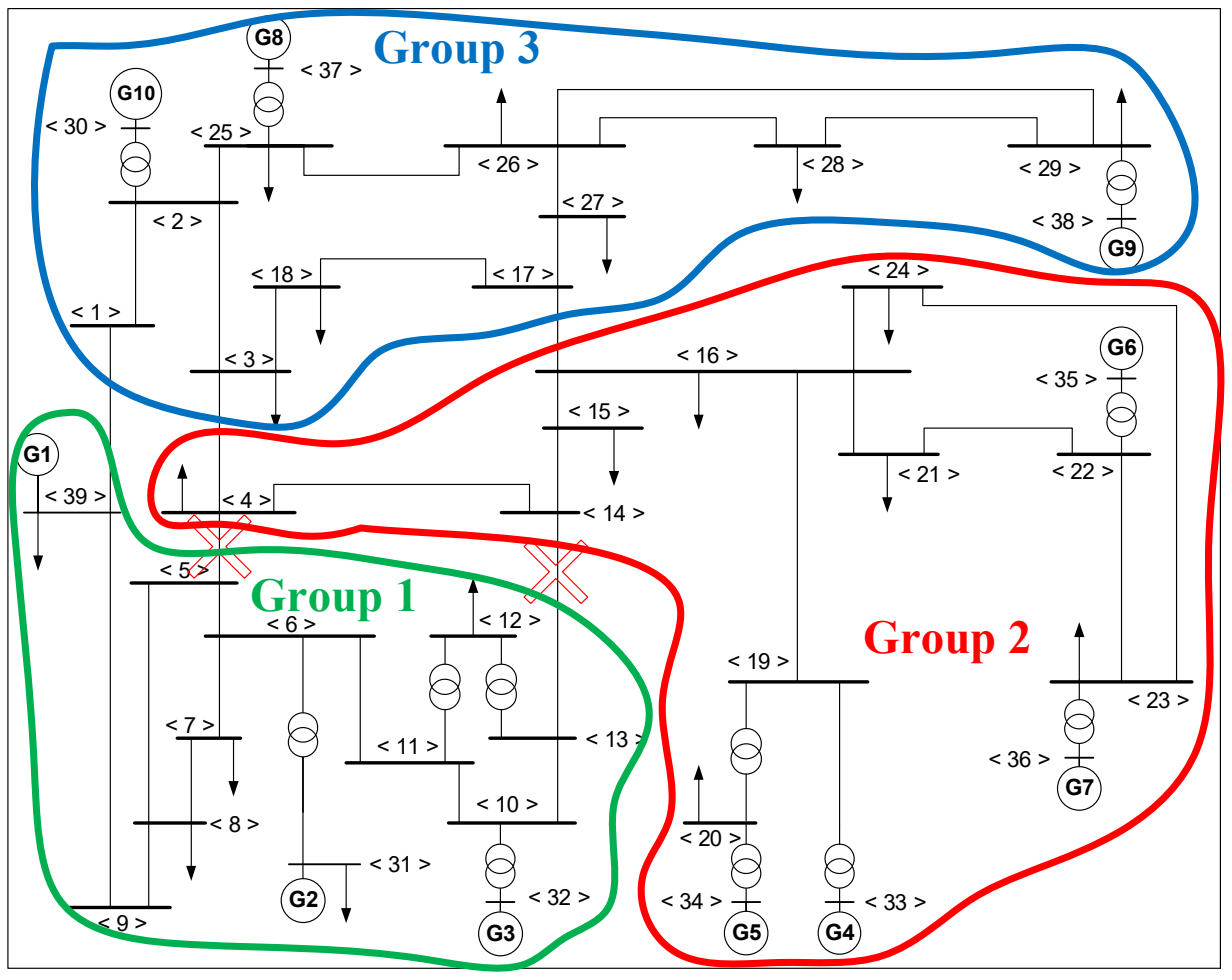

Fig. 1. IEEE 39-bus system

| | 4 | 14 | 15 | 16 | 19 | 20 | 21 | 22 | 23 | 24 | 33 | 34 | 35 | 36 |
|---|---|---|---|---|---|---|---|---|---|---|---|---|---|---|
| 4 | | 1454 | 0 | 0 | 0 | 0 | 0 | 0 | 0 | 0 | 0 | 0 | 0 | 0 |
| 14 | 1454 | | 1784 | 0 | 0 | 0 | 0 | 0 | 0 | 0 | 0 | 0 | 0 | 0 |
| 15 | 0 | 1784 | | 2638 | 0 | 0 | 0 | 0 | 0 | 0 | 0 | 0 | 0 | 0 |
| 16 | 0 | 0 | 2638 | | 2546 | 0 | 917 | 0 | 0 | 445 | 0 | 0 | 0 | 0 |
| 19 | 0 | 0 | 0 | 2546 | | 903 | 0 | 0 | 0 | 0 | 1881 | 0 | 0 | 0 |
| 20 | 0 | 0 | 0 | 0 | 903 | | 0 | 0 | 0 | 0 | 0 | 2782 | 0 | 0 |
| 21 | 0 | 0 | 0 | 917 | 0 | 0 | | 1733 | 0 | 0 | 0 | 0 | 0 | 0 |
| 22 | 0 | 0 | 0 | 0 | 0 | 0 | 1733 | | 129 | 0 | 0 | 0 | 1967 | 0 |
| 23 | 0 | 0 | 0 | 0 | 0 | 0 | 0 | 129 | | 983 | 0 | 0 | 0 | 1709 |
| 24 | 0 | 0 | 0 | 445 | 0 | 0 | 0 | 0 | 983 | | 0 | 0 | 0 | 0 |
| 33 | 0 | 0 | 0 | 0 | 1881 | 0 | 0 | 0 | 0 | 0 | | 0 | 0 | 0 |
| 34 | 0 | 0 | 0 | 0 | 0 | 2782 | 0 | 0 | 0 | 0 | 0 | | 0 | 0 |
| 35 | 0 | 0 | 0 | 0 | 0 | 0 | 0 | 1967 | 0 | 0 | 0 | 0 | | 0 |
| 36 | 0 | 0 | 0 | 0 | 0 | 0 | 0 | 0 | 1709 | 0 | 0 | 0 | 0 | |
| 1 | 0 | 0 | 0 | 0 | 0 | 0 | 0 | 0 | 0 | 0 | 0 | 0 | 0 | 0 |
| 5 | 496 | 0 | 0 | 0 | 0 | 0 | 0 | 0 | 0 | 0 | 0 | 0 | 0 | 0 |
| 6 | 0 | 0 | 0 | 0 | 0 | 0 | 0 | 0 | 0 | 0 | 0 | 0 | 0 | 0 |
| 7 | 0 | 0 | 0 | 0 | 0 | 0 | 0 | 0 | 0 | 0 | 0 | 0 | 0 | 0 |
| 8 | 0 | 0 | 0 | 0 | 0 | 0 | 0 | 0 | 0 | 0 | 0 | 0 | 0 | 0 |
| 9 | 0 | 0 | 0 | 0 | 0 | 0 | 0 | 0 | 0 | 0 | 0 | 0 | 0 | 0 |
| 10 | 0 | 0 | 0 | 0 | 0 | 0 | 0 | 0 | 0 | 0 | 0 | 0 | 0 | 0 |
| 11 | 0 | 0 | 0 | 0 | 0 | 0 | 0 | 0 | 0 | 0 | 0 | 0 | 0 | 0 |
| 12 | 0 | 0 | 0 | 0 | 0 | 0 | 0 | 0 | 0 | 0 | 0 | 0 | 0 | 0 |
| 13 | 0 | 713 | 0 | 0 | 0 | 0 | 0 | 0 | 0 | 0 | 0 | 0 | 0 | 0 |
| 31 | 0 | 0 | 0 | 0 | 0 | 0 | 0 | 0 | 0 | 0 | 0 | 0 | 0 | 0 |
| 32 | 0 | 0 | 0 | 0 | 0 | 0 | 0 | 0 | 0 | 0 | 0 | 0 | 0 | 0 |
| 39 | 0 | 0 | 0 | 0 | 0 | 0 | 0 | 0 | 0 | 0 | 0 | 0 | 0 | 0 |
| 2 | 0 | 0 | 0 | 0 | 0 | 0 | 0 | 0 | 0 | 0 | 0 | 0 | 0 | 0 |
| 3 | 0 | 0 | 0 | 0 | 0 | 0 | 0 | 0 | 0 | 0 | 0 | 0 | 0 | 0 |
| 17 | 0 | 0 | 0 | 0 | 0 | 0 | 0 | 0 | 0 | 0 | 0 | 0 | 0 | 0 |
| 18 | 0 | 0 | 0 | 0 | 0 | 0 | 0 | 0 | 0 | 0 | 0 | 0 | 0 | 0 |
| 25 | 0 | 0 | 0 | 0 | 0 | 0 | 0 | 0 | 0 | 0 | 0 | 0 | 0 | 0 |
| 26 | 0 | 0 | 0 | 0 | 0 | 0 | 0 | 0 | 0 | 0 | 0 | 0 | 0 | 0 |
| 27 | 0 | 0 | 0 | 0 | 0 | 0 | 0 | 0 | 0 | 0 | 0 | 0 | 0 | 0 |
| 28 | 0 | 0 | 0 | 0 | 0 | 0 | 0 | 0 | 0 | 0 | 0 | 0 | 0 | 0 |
| 29 | 0 | 0 | 0 | 0 | 0 | 0 | 0 | 0 | 0 | 0 | 0 | 0 | 0 | 0 |
| 30 | 0 | 0 | 0 | 0 | 0 | 0 | 0 | 0 | 0 | 0 | 0 | 0 | 0 | 0 |
| 37 | 0 | 0 | 0 | 0 | 0 | 0 | 0 | 0 | 0 | 0 | 0 | 0 | 0 | 0 |
| 38 | 0 | 0 | 0 | 0 | 0 | 0 | 0 | 0 | 0 | 0 | 0 | 0 | 0 | 0 |

| | 1 | 5 | 6 | 7 | 8 | 9 | 10 | 11 | 12 | 13 | 31 | 32 | 39 |
|---|---|---|---|---|---|---|---|---|---|---|---|---|---|
| 4 | 0 | 496 | 0 | 0 | 0 | 0 | 0 | 0 | 0 | 0 | 0 | 0 | 0 |
| 14 | 0 | 0 | 0 | 0 | 0 | 0 | 0 | 0 | 0 | 713 | 0 | 0 | 0 |
| 15 | 0 | 0 | 0 | 0 | 0 | 0 | 0 | 0 | 0 | 0 | 0 | 0 | 0 |
| 16 | 0 | 0 | 0 | 0 | 0 | 0 | 0 | 0 | 0 | 0 | 0 | 0 | 0 |
| 19 | 0 | 0 | 0 | 0 | 0 | 0 | 0 | 0 | 0 | 0 | 0 | 0 | 0 |
| 20 | 0 | 0 | 0 | 0 | 0 | 0 | 0 | 0 | 0 | 0 | 0 | 0 | 0 |
| 21 | 0 | 0 | 0 | 0 | 0 | 0 | 0 | 0 | 0 | 0 | 0 | 0 | 0 |
| 22 | 0 | 0 | 0 | 0 | 0 | 0 | 0 | 0 | 0 | 0 | 0 | 0 | 0 |
| 23 | 0 | 0 | 0 | 0 | 0 | 0 | 0 | 0 | 0 | 0 | 0 | 0 | 0 |
| 24 | 0 | 0 | 0 | 0 | 0 | 0 | 0 | 0 | 0 | 0 | 0 | 0 | 0 |
| 33 | 0 | 0 | 0 | 0 | 0 | 0 | 0 | 0 | 0 | 0 | 0 | 0 | 0 |
| 34 | 0 | 0 | 0 | 0 | 0 | 0 | 0 | 0 | 0 | 0 | 0 | 0 | 0 |
| 35 | 0 | 0 | 0 | 0 | 0 | 0 | 0 | 0 | 0 | 0 | 0 | 0 | 0 |
| 36 | 0 | 0 | 0 | 0 | 0 | 0 | 0 | 0 | 0 | 0 | 0 | 0 | 0 |
| 1 | | 0 | 0 | 0 | 0 | 0 | 0 | 0 | 0 | 0 | 0 | 0 | 1059 |
| 5 | 0 | | 1237 | 0 | 1256 | 0 | 0 | 0 | 0 | 0 | 0 | 0 | 0 |
| 6 | 0 | 1237 | | 1455 | 0 | 0 | 0 | 1432 | 0 | 0 | 1626 | 0 | 0 |
| 7 | 0 | 0 | 1455 | | 816 | 0 | 0 | 0 | 0 | 0 | 0 | 0 | 0 |
| 8 | 0 | 1256 | 0 | 816 | | 778 | 0 | 0 | 0 | 0 | 0 | 0 | 0 |
| 9 | 0 | 0 | 0 | 0 | 778 | | 0 | 0 | 0 | 0 | 0 | 0 | 776 |
| 10 | 0 | 0 | 0 | 0 | 0 | 0 | | 1405 | 0 | 780 | 0 | 1774 | 0 |
| 11 | 0 | 0 | 1432 | 0 | 0 | 0 | 1405 | | 135 | 0 | 0 | 0 | 0 |
| 12 | 0 | 0 | 0 | 0 | 0 | 0 | 0 | 135 | | 117 | 0 | 0 | 0 |
| 13 | 0 | 0 | 0 | 0 | 0 | 0 | 780 | 0 | 117 | | 0 | 0 | 0 |
| 31 | 0 | 0 | 1626 | 0 | 0 | 0 | 0 | 0 | 0 | 0 | | 0 | 0 |
| 32 | 0 | 0 | 0 | 0 | 0 | 0 | 1774 | 0 | 0 | 0 | 0 | | 0 |
| 39 | 1059 | 0 | 0 | 0 | 0 | 776 | 0 | 0 | 0 | 0 | 0 | 0 | |
| 2 | 1078 | 0 | 0 | 0 | 0 | 0 | 0 | 0 | 0 | 0 | 0 | 0 | 0 |
| 3 | 0 | 0 | 0 | 0 | 0 | 0 | 0 | 0 | 0 | 0 | 0 | 0 | 0 |
| 17 | 0 | 0 | 0 | 0 | 0 | 0 | 0 | 0 | 0 | 0 | 0 | 0 | 0 |
| 18 | 0 | 0 | 0 | 0 | 0 | 0 | 0 | 0 | 0 | 0 | 0 | 0 | 0 |
| 25 | 0 | 0 | 0 | 0 | 0 | 0 | 0 | 0 | 0 | 0 | 0 | 0 | 0 |
| 26 | 0 | 0 | 0 | 0 | 0 | 0 | 0 | 0 | 0 | 0 | 0 | 0 | 0 |
| 27 | 0 | 0 | 0 | 0 | 0 | 0 | 0 | 0 | 0 | 0 | 0 | 0 | 0 |
| 28 | 0 | 0 | 0 | 0 | 0 | 0 | 0 | 0 | 0 | 0 | 0 | 0 | 0 |
| 29 | 0 | 0 | 0 | 0 | 0 | 0 | 0 | 0 | 0 | 0 | 0 | 0 | 0 |
| 30 | 0 | 0 | 0 | 0 | 0 | 0 | 0 | 0 | 0 | 0 | 0 | 0 | 0 |
| 37 | 0 | 0 | 0 | 0 | 0 | 0 | 0 | 0 | 0 | 0 | 0 | 0 | 0 |
| 38 | 0 | 0 | 0 | 0 | 0 | 0 | 0 | 0 | 0 | 0 | 0 | 0 | 0 |

| | 2 | 3 | 17 | 18 | 25 | 26 | 27 | 28 | 29 | 30 | 37 | 38 |
|---|---|---|---|---|---|---|---|---|---|---|---|---|
| 4 | 0 | 0 | 0 | 0 | 0 | 0 | 0 | 0 | 0 | 0 | 0 | 0 |
| 14 | 0 | 0 | 0 | 0 | 0 | 0 | 0 | 0 | 0 | 0 | 0 | 0 |
| 15 | 0 | 0 | 0 | 0 | 0 | 0 | 0 | 0 | 0 | 0 | 0 | 0 |
| 16 | 0 | 0 | 0 | 0 | 0 | 0 | 0 | 0 | 0 | 0 | 0 | 0 |
| 19 | 0 | 0 | 0 | 0 | 0 | 0 | 0 | 0 | 0 | 0 | 0 | 0 |
| 20 | 0 | 0 | 0 | 0 | 0 | 0 | 0 | 0 | 0 | 0 | 0 | 0 |
| 21 | 0 | 0 | 0 | 0 | 0 | 0 | 0 | 0 | 0 | 0 | 0 | 0 |
| 22 | 0 | 0 | 0 | 0 | 0 | 0 | 0 | 0 | 0 | 0 | 0 | 0 |
| 23 | 0 | 0 | 0 | 0 | 0 | 0 | 0 | 0 | 0 | 0 | 0 | 0 |
| 24 | 0 | 0 | 0 | 0 | 0 | 0 | 0 | 0 | 0 | 0 | 0 | 0 |
| 33 | 0 | 0 | 0 | 0 | 0 | 0 | 0 | 0 | 0 | 0 | 0 | 0 |
| 34 | 0 | 0 | 0 | 0 | 0 | 0 | 0 | 0 | 0 | 0 | 0 | 0 |
| 35 | 0 | 0 | 0 | 0 | 0 | 0 | 0 | 0 | 0 | 0 | 0 | 0 |
| 36 | 0 | 0 | 0 | 0 | 0 | 0 | 0 | 0 | 0 | 0 | 0 | 0 |
| 1 | 1078 | 0 | 0 | 0 | 0 | 0 | 0 | 0 | 0 | 0 | 0 | 0 |
| 5 | 0 | 0 | 0 | 0 | 0 | 0 | 0 | 0 | 0 | 0 | 0 | 0 |
| 6 | 0 | 0 | 0 | 0 | 0 | 0 | 0 | 0 | 0 | 0 | 0 | 0 |
| 7 | 0 | 0 | 0 | 0 | 0 | 0 | 0 | 0 | 0 | 0 | 0 | 0 |
| 8 | 0 | 0 | 0 | 0 | 0 | 0 | 0 | 0 | 0 | 0 | 0 | 0 |
| 9 | 0 | 0 | 0 | 0 | 0 | 0 | 0 | 0 | 0 | 0 | 0 | 0 |
| 10 | 0 | 0 | 0 | 0 | 0 | 0 | 0 | 0 | 0 | 0 | 0 | 0 |
| 11 | 0 | 0 | 0 | 0 | 0 | 0 | 0 | 0 | 0 | 0 | 0 | 0 |
| 12 | 0 | 0 | 0 | 0 | 0 | 0 | 0 | 0 | 0 | 0 | 0 | 0 |
| 13 | 0 | 0 | 0 | 0 | 0 | 0 | 0 | 0 | 0 | 0 | 0 | 0 |
| 31 | 0 | 0 | 0 | 0 | 0 | 0 | 0 | 0 | 0 | 0 | 0 | 0 |
| 32 | 0 | 0 | 0 | 0 | 0 | 0 | 0 | 0 | 0 | 0 | 0 | 0 |
| 39 | 0 | 0 | 0 | 0 | 0 | 0 | 0 | 0 | 0 | 0 | 0 | 0 |
| 2 | | 1662 | 0 | 0 | 215 | 0 | 0 | 0 | 0 | 1094 | 0 | 0 |
| 3 | 1662 | | 0 | 663 | 0 | 0 | 0 | 0 | 0 | 0 | 0 | 0 |
| 17 | 0 | 0 | | 185 | 0 | 0 | 182 | 0 | 0 | 0 | 0 | 0 |
| 18 | 0 | 663 | 185 | | 0 | 0 | 0 | 0 | 0 | 0 | 0 | 0 |
| 25 | 215 | 0 | 0 | 0 | | 830 | 0 | 0 | 0 | 0 | 1332 | 0 |
| 26 | 0 | 0 | 0 | 0 | 830 | | 704 | 112 | 244 | 0 | 0 | 0 |
| 27 | 0 | 0 | 182 | 0 | 0 | 704 | | 0 | 0 | 0 | 0 | 0 |
| 28 | 0 | 0 | 0 | 0 | 0 | 112 | 0 | | 697 | 0 | 0 | 0 |
| 29 | 0 | 0 | 0 | 0 | 0 | 244 | 0 | 697 | | 0 | 0 | 1801 |
| 30 | 1094 | 0 | 0 | 0 | 0 | 0 | 0 | 0 | 0 | | 0 | 0 |
| 37 | 0 | 0 | 0 | 0 | 1332 | 0 | 0 | 0 | 0 | 0 | | 0 |
| 38 | 0 | 0 | 0 | 0 | 0 | 0 | 0 | 0 | 1801 | 0 | 0 | |

Fig. 2. Matrix of apparent power at t=4.15 sec

Two short circuits (SC) events occurred in lines 3-4 and 16-17 at t= 2 sec and t=3 sec respectively. The faults are cleared after 400 msec by opening the line switches from the substations, while the simulation lasts for t=8 sec. Fig. 3, Fig. 4, and Fig. 5 demonstrate the generator's rotor angle, voltage angle of buses, and the system frequency respectively, that indicate the system instability following the second event. The proposed solution approach is applied to the system to determine the islanding boundaries; the quality of each island is then evaluated by calculating the dynamic behavior and the power mismatch in the islands. It can be observed in Fig. 3, that if no control action is undertaken, the system loses synchronism at about 4.25 sec. Indeed, real-time simulation in DIgSILENT indicates out of step at t=4.25 sec for generators. As noticed, the system is divided into three groups, which are not balanced. The frequency of the generators and the loss of synchronism are a clear indication that the system should be split.

To proceed with the separation methodology, we consider t=4.15 sec, 100 msec before the loose of synchronism in the system, as the point when the system separation is applied. To this end, the Ks matrix [13] is first obtained at t=4.15, and the modularity clustering approach is employed to cluster Ks Matrix into the coherent groups of generators. As can be seen in Fig. 7, the generators are divided into three coherent groups of {G1, G2, G3}, {G4, G5, G6, G7}, {G8, G9, G10}. To have a stable island, we need to have strong coherency among generators on the island. Accordingly, the coherent generators must be allocated together on one island. This can be done by the "must-link" concept in the constraint clustering approach. In other words, coherent generators must be linked together on one island.

According to the proposed approach, the matrix of apparent power is clustered to the number of coherent groups in such a way that coherent generators being in the same group. Fig. 2 shows the partitioned apparent power matrix at t=4.15 sec. As can be seen, the network is partitioned into three groups, in which the cuts promise lowest power mismatch in groups.

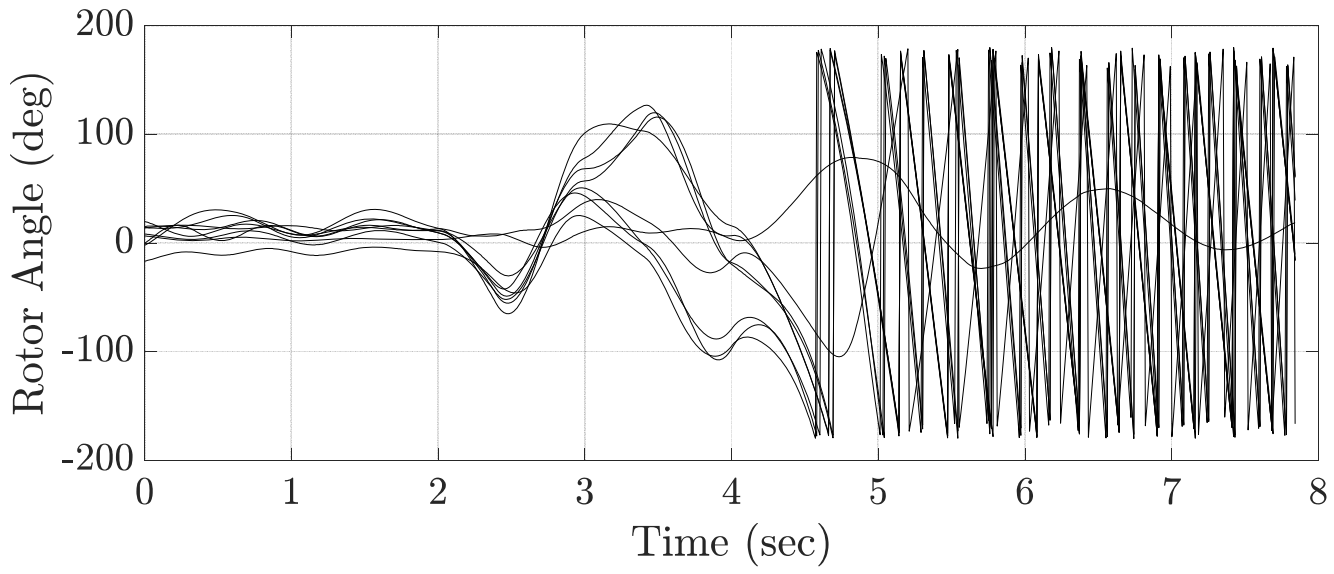

Fig. 3. Generator rotor angles in Case I

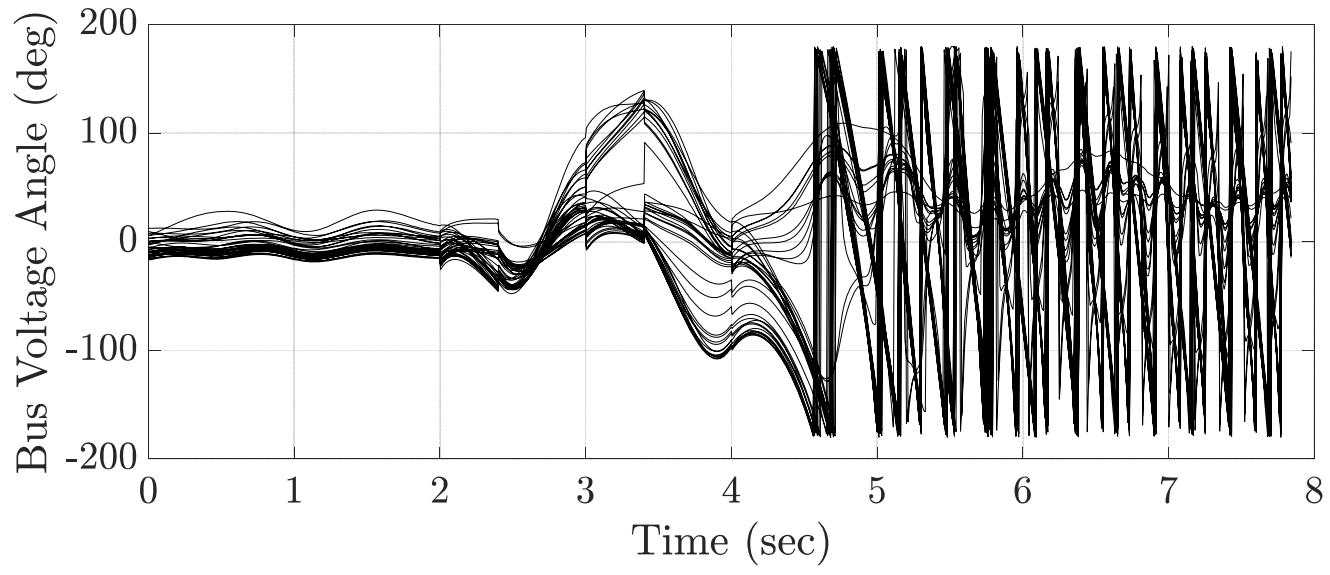

Fig. 4. Bus voltage angles in Case I

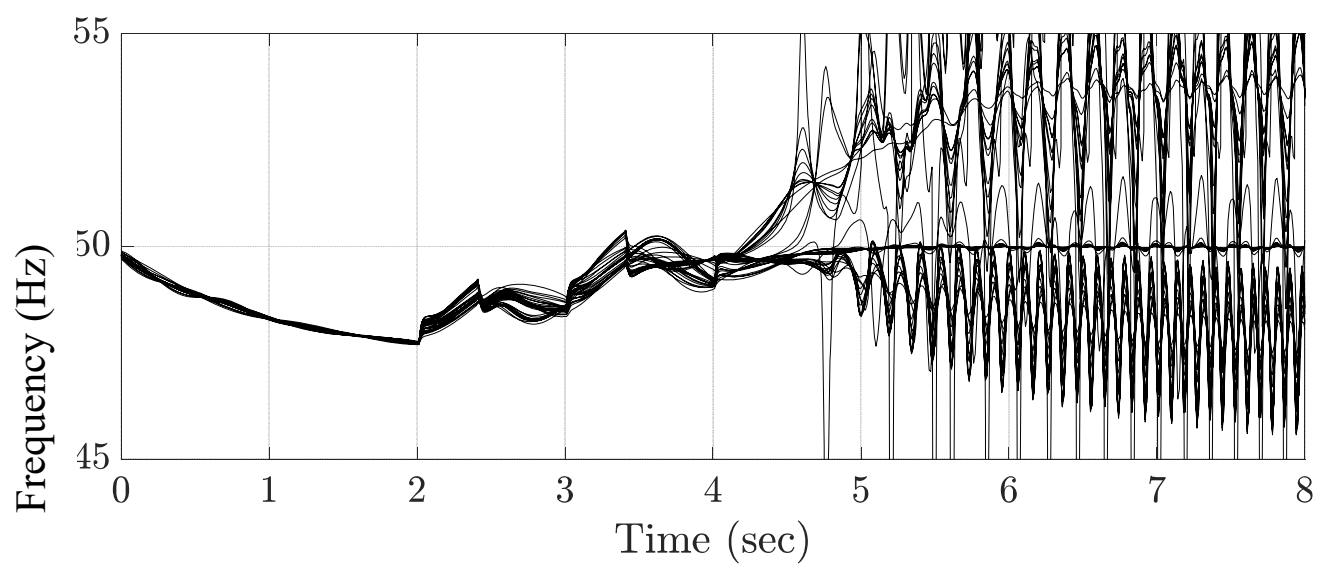

Fig. 5. Bus frequencies in Case I

| | 4 | 14 | 15 | 16 | 19 | 20 | 21 | 22 | 23 | 24 | 33 | 34 | 35 | 36 | 1 | 5 | 6 | 7 | 8 | 9 | 10 | 11 | 12 | 13 | 31 | 32 | 39 | 2 | 3 | 17 | 18 | 25 | 26 | 27 | 28 | 29 | 30 | 37 | 38 |
|---|---|---|---|---|---|---|---|---|---|---|---|---|---|---|---|---|---|---|---|---|---|---|---|---|---|---|---|---|---|---|---|---|---|---|---|---|---|---|---|
| 4 | | 1.00 | 0.97 | 0.95 | 0.98 | 1.00 | 0.91 | 0.85 | 0.85 | 0.95 | 0.98 | 1.00 | 0.81 | 0.73 | 0.76 | 1.00 | 1.00 | 1.00 | 1.00 | 0.94 | 1.00 | 1.00 | 1.00 | 1.00 | 0.99 | 0.98 | 0.86 | 0.58 | 0.57 | 0.55 | 0.56 | 0.55 | 0.53 | 0.54 | 0.50 | 0.50 | 0.57 | 0.50 | 0.48 |
| 14 | 1.00 | | 0.99 | 0.98 | 0.99 | 1.00 | 0.95 | 0.90 | 0.89 | 0.97 | 0.99 | 1.00 | 0.86 | 0.79 | 0.82 | 1.00 | 1.00 | 1.00 | 1.00 | 0.97 | 1.00 | 1.00 | 1.00 | 1.00 | 0.98 | 0.99 | 0.91 | 0.65 | 0.64 | 0.63 | 0.64 | 0.63 | 0.61 | 0.62 | 0.58 | 0.58 | 0.64 | 0.58 | 0.56 |
| 15 | 0.97 | 0.99 | | 1.00 | 1.00 | 0.99 | 0.98 | 0.95 | 0.95 | 1.00 | 1.00 | 0.97 | 0.92 | 0.87 | 0.89 | 0.97 | 0.98 | 0.97 | 0.98 | 0.99 | 0.99 | 0.98 | 0.99 | 0.99 | 0.94 | 1.00 | 0.95 | 0.75 | 0.74 | 0.73 | 0.74 | 0.73 | 0.71 | 0.72 | 0.69 | 0.68 | 0.74 | 0.69 | 0.67 |
| 16 | 0.95 | 0.98 | 1.00 | | 0.99 | 0.98 | 0.99 | 0.97 | 0.97 | 1.00 | 1.00 | 0.96 | 0.95 | 0.90 | 0.92 | 0.96 | 0.96 | 0.95 | 0.96 | 1.00 | 0.98 | 0.97 | 0.97 | 0.98 | 0.91 | 0.99 | 0.97 | 0.80 | 0.79 | 0.78 | 0.78 | 0.77 | 0.76 | 0.77 | 0.74 | 0.73 | 0.79 | 0.74 | 0.72 |
| 19 | 0.98 | 0.99 | 1.00 | 0.99 | | 0.99 | 0.98 | 0.94 | 0.93 | 0.99 | 1.00 | 0.98 | 0.91 | 0.85 | 0.87 | 0.98 | 0.98 | 0.98 | 0.98 | 0.99 | 0.99 | 0.99 | 0.99 | 0.99 | 0.95 | 1.00 | 0.94 | 0.73 | 0.72 | 0.71 | 0.71 | 0.70 | 0.69 | 0.70 | 0.66 | 0.66 | 0.72 | 0.66 | 0.64 |
| 20 | 1.00 | 1.00 | 0.99 | 0.98 | 0.99 | | 0.95 | 0.90 | 0.89 | 0.97 | 0.99 | 1.00 | 0.86 | 0.79 | 0.81 | 1.00 | 1.00 | 1.00 | 1.00 | 0.97 | 1.00 | 1.00 | 1.00 | 1.00 | 0.98 | 0.99 | 0.90 | 0.65 | 0.64 | 0.63 | 0.63 | 0.62 | 0.61 | 0.62 | 0.58 | 0.57 | 0.64 | 0.58 | 0.56 |
| 21 | 0.91 | 0.95 | 0.98 | 0.99 | 0.98 | 0.95 | | 0.99 | 0.99 | 1.00 | 0.98 | 0.91 | 0.98 | 0.94 | 0.96 | 0.92 | 0.92 | 0.91 | 0.92 | 1.00 | 0.94 | 0.94 | 0.94 | 0.94 | 0.86 | 0.98 | 0.99 | 0.86 | 0.85 | 0.84 | 0.85 | 0.84 | 0.83 | 0.84 | 0.81 | 0.81 | 0.85 | 0.81 | 0.79 |
| 22 | 0.85 | 0.90 | 0.95 | 0.97 | 0.94 | 0.90 | 0.99 | | 1.00 | 0.98 | 0.94 | 0.85 | 1.00 | 0.98 | 0.99 | 0.86 | 0.86 | 0.85 | 0.86 | 0.98 | 0.89 | 0.88 | 0.89 | 0.89 | 0.79 | 0.94 | 1.00 | 0.92 | 0.91 | 0.91 | 0.91 | 0.90 | 0.90 | 0.90 | 0.88 | 0.88 | 0.91 | 0.88 | 0.87 |
| 23 | 0.85 | 0.89 | 0.95 | 0.97 | 0.93 | 0.89 | 0.99 | 1.00 | | 0.97 | 0.94 | 0.85 | 1.00 | 0.98 | 0.99 | 0.85 | 0.85 | 0.85 | 0.85 | 0.97 | 0.89 | 0.88 | 0.88 | 0.89 | 0.78 | 0.93 | 1.00 | 0.92 | 0.92 | 0.91 | 0.92 | 0.91 | 0.90 | 0.91 | 0.89 | 0.88 | 0.92 | 0.89 | 0.87 |
| 24 | 0.95 | 0.97 | 1.00 | 1.00 | 0.99 | 0.97 | 1.00 | 0.98 | 0.97 | | 0.99 | 0.95 | 0.95 | 0.91 | 0.93 | 0.95 | 0.95 | 0.95 | 0.95 | 1.00 | 0.97 | 0.96 | 0.97 | 0.97 | 0.90 | 0.99 | 0.98 | 0.81 | 0.80 | 0.79 | 0.80 | 0.79 | 0.78 | 0.78 | 0.76 | 0.75 | 0.80 | 0.75 | 0.74 |
| 33 | 0.98 | 0.99 | 1.00 | 1.00 | 1.00 | 0.99 | 0.98 | 0.94 | 0.94 | 0.99 | | 0.98 | 0.91 | 0.86 | 0.88 | 0.98 | 0.98 | 0.98 | 0.98 | 0.99 | 0.99 | 0.99 | 0.99 | 0.99 | 0.95 | 1.00 | 0.95 | 0.74 | 0.73 | 0.72 | 0.72 | 0.71 | 0.70 | 0.71 | 0.68 | 0.67 | 0.73 | 0.67 | 0.65 |
| 34 | 1.00 | 1.00 | 0.97 | 0.96 | 0.98 | 1.00 | 0.91 | 0.85 | 0.85 | 0.95 | 0.98 | | 0.81 | 0.73 | 0.76 | 1.00 | 1.00 | 1.00 | 1.00 | 0.94 | 1.00 | 1.00 | 1.00 | 1.00 | 0.99 | 0.98 | 0.86 | 0.58 | 0.57 | 0.56 | 0.56 | 0.55 | 0.53 | 0.54 | 0.51 | 0.50 | 0.57 | 0.50 | 0.48 |
| 35 | 0.81 | 0.86 | 0.92 | 0.95 | 0.91 | 0.86 | 0.98 | 1.00 | 1.00 | 0.95 | 0.91 | 0.81 | | 0.99 | 1.00 | 0.81 | 0.81 | 0.81 | 0.81 | 0.96 | 0.85 | 0.84 | 0.85 | 0.85 | 0.73 | 0.91 | 0.99 | 0.95 | 0.95 | 0.94 | 0.94 | 0.94 | 0.93 | 0.93 | 0.92 | 0.91 | 0.94 | 0.92 | 0.90 |
| 36 | 0.73 | 0.79 | 0.87 | 0.90 | 0.85 | 0.79 | 0.94 | 0.98 | 0.98 | 0.91 | 0.86 | 0.73 | 0.99 | | 1.00 | 0.73 | 0.74 | 0.73 | 0.73 | 0.91 | 0.78 | 0.77 | 0.77 | 0.78 | 0.65 | 0.85 | 0.97 | 0.98 | 0.98 | 0.97 | 0.98 | 0.97 | 0.97 | 0.97 | 0.96 | 0.96 | 0.98 | 0.96 | 0.95 |
| 1 | 0.76 | 0.82 | 0.89 | 0.92 | 0.87 | 0.81 | 0.96 | 0.99 | 0.99 | 0.93 | 0.88 | 0.76 | 1.00 | 1.00 | | 0.77 | 0.77 | 0.76 | 0.77 | 0.93 | 0.81 | 0.80 | 0.80 | 0.81 | 0.68 | 0.87 | 0.98 | 0.97 | 0.97 | 0.96 | 0.96 | 0.96 | 0.95 | 0.96 | 0.94 | 0.94 | 0.97 | 0.94 | 0.93 |
| 5 | 1.00 | 1.00 | 0.97 | 0.96 | 0.98 | 1.00 | 0.92 | 0.86 | 0.85 | 0.95 | 0.98 | 1.00 | 0.81 | 0.73 | 0.77 | | 1.00 | 1.00 | 1.00 | 0.94 | 1.00 | 1.00 | 1.00 | 1.00 | 0.99 | 0.98 | 0.87 | 0.59 | 0.58 | 0.56 | 0.57 | 0.56 | 0.54 | 0.55 | 0.51 | 0.50 | 0.57 | 0.51 | 0.49 |
| 6 | 1.00 | 1.00 | 0.98 | 0.96 | 0.98 | 1.00 | 0.92 | 0.86 | 0.85 | 0.95 | 0.98 | 1.00 | 0.81 | 0.74 | 0.77 | 1.00 | | 1.00 | 1.00 | 0.95 | 1.00 | 1.00 | 1.00 | 1.00 | 0.99 | 0.98 | 0.87 | 0.59 | 0.58 | 0.56 | 0.57 | 0.56 | 0.54 | 0.55 | 0.51 | 0.51 | 0.58 | 0.51 | 0.49 |
| 7 | 1.00 | 1.00 | 0.97 | 0.95 | 0.98 | 1.00 | 0.91 | 0.85 | 0.85 | 0.95 | 0.98 | 1.00 | 0.81 | 0.73 | 0.76 | 1.00 | 1.00 | | 1.00 | 0.94 | 1.00 | 1.00 | 1.00 | 1.00 | 0.99 | 0.98 | 0.86 | 0.58 | 0.57 | 0.56 | 0.56 | 0.55 | 0.53 | 0.54 | 0.51 | 0.50 | 0.57 | 0.50 | 0.48 |
| 8 | 1.00 | 1.00 | 0.98 | 0.96 | 0.98 | 1.00 | 0.92 | 0.86 | 0.85 | 0.95 | 0.98 | 1.00 | 0.81 | 0.73 | 0.77 | 1.00 | 1.00 | 1.00 | | 0.95 | 1.00 | 1.00 | 1.00 | 1.00 | 0.99 | 0.98 | 0.87 | 0.59 | 0.58 | 0.56 | 0.57 | 0.56 | 0.54 | 0.55 | 0.51 | 0.51 | 0.58 | 0.51 | 0.49 |
| 9 | 0.94 | 0.97 | 0.99 | 1.00 | 0.99 | 0.97 | 1.00 | 0.98 | 0.97 | 1.00 | 0.99 | 0.94 | 0.96 | 0.91 | 0.93 | 0.94 | 0.95 | 0.94 | 0.95 | | 0.97 | 0.96 | 0.96 | 0.97 | 0.90 | 0.99 | 0.98 | 0.82 | 0.81 | 0.80 | 0.81 | 0.80 | 0.78 | 0.79 | 0.76 | 0.76 | 0.81 | 0.76 | 0.74 |
| 10 | 1.00 | 1.00 | 0.99 | 0.98 | 0.99 | 1.00 | 0.94 | 0.89 | 0.89 | 0.97 | 0.99 | 1.00 | 0.85 | 0.78 | 0.81 | 1.00 | 1.00 | 1.00 | 1.00 | 0.97 | | 1.00 | 1.00 | 1.00 | 0.98 | 0.99 | 0.90 | 0.65 | 0.64 | 0.62 | 0.63 | 0.62 | 0.60 | 0.61 | 0.57 | 0.57 | 0.63 | 0.57 | 0.55 |
| 11 | 1.00 | 1.00 | 0.98 | 0.97 | 0.99 | 1.00 | 0.94 | 0.88 | 0.88 | 0.96 | 0.99 | 1.00 | 0.84 | 0.77 | 0.80 | 1.00 | 1.00 | 1.00 | 1.00 | 0.96 | 1.00 | | 1.00 | 1.00 | 0.99 | 0.99 | 0.89 | 0.63 | 0.62 | 0.60 | 0.61 | 0.60 | 0.58 | 0.59 | 0.55 | 0.55 | 0.61 | 0.55 | 0.53 |
| 12 | 1.00 | 1.00 | 0.99 | 0.97 | 0.99 | 1.00 | 0.94 | 0.89 | 0.88 | 0.97 | 0.99 | 1.00 | 0.85 | 0.77 | 0.80 | 1.00 | 1.00 | 1.00 | 1.00 | 0.96 | 1.00 | 1.00 | | 1.00 | 0.98 | 0.99 | 0.90 | 0.64 | 0.63 | 0.61 | 0.62 | 0.61 | 0.59 | 0.60 | 0.56 | 0.56 | 0.62 | 0.56 | 0.54 |
| 13 | 1.00 | 1.00 | 0.99 | 0.98 | 0.99 | 1.00 | 0.94 | 0.89 | 0.89 | 0.97 | 0.99 | 1.00 | 0.85 | 0.78 | 0.81 | 1.00 | 1.00 | 1.00 | 1.00 | 0.97 | 1.00 | 1.00 | 1.00 | | 0.98 | 0.99 | 0.90 | 0.65 | 0.64 | 0.62 | 0.63 | 0.62 | 0.60 | 0.61 | 0.58 | 0.57 | 0.63 | 0.57 | 0.55 |
| 31 | 0.99 | 0.98 | 0.94 | 0.91 | 0.95 | 0.98 | 0.86 | 0.79 | 0.78 | 0.90 | 0.95 | 0.99 | 0.73 | 0.65 | 0.68 | 0.99 | 0.99 | 0.99 | 0.99 | 0.90 | 0.98 | 0.99 | 0.98 | 0.98 | | 0.95 | 0.80 | 0.49 | 0.47 | 0.46 | 0.46 | 0.45 | 0.43 | 0.44 | 0.40 | 0.40 | 0.47 | 0.40 | 0.38 |
| 32 | 0.98 | 0.99 | 1.00 | 0.99 | 1.00 | 0.99 | 0.98 | 0.94 | 0.93 | 0.99 | 1.00 | 0.98 | 0.91 | 0.85 | 0.87 | 0.98 | 0.98 | 0.98 | 0.98 | 0.99 | 0.99 | 0.99 | 0.99 | 0.99 | 0.95 | | 0.94 | 0.73 | 0.72 | 0.71 | 0.71 | 0.70 | 0.69 | 0.70 | 0.66 | 0.66 | 0.72 | 0.66 | 0.64 |
| 39 | 0.86 | 0.91 | 0.95 | 0.97 | 0.94 | 0.90 | 0.99 | 1.00 | 1.00 | 0.98 | 0.95 | 0.86 | 0.99 | 0.97 | 0.98 | 0.87 | 0.87 | 0.86 | 0.87 | 0.98 | 0.90 | 0.89 | 0.90 | 0.90 | 0.80 | 0.94 | | 0.91 | 0.91 | 0.90 | 0.90 | 0.90 | 0.89 | 0.89 | 0.87 | 0.86 | 0.90 | 0.87 | 0.85 |
| 2 | 0.58 | 0.65 | 0.75 | 0.80 | 0.73 | 0.65 | 0.86 | 0.92 | 0.92 | 0.81 | 0.74 | 0.58 | 0.95 | 0.98 | 0.97 | 0.59 | 0.59 | 0.58 | 0.59 | 0.82 | 0.65 | 0.63 | 0.64 | 0.65 | 0.49 | 0.73 | 0.91 | | 1.00 | 1.00 | 1.00 | 1.00 | 1.00 | 1.00 | 1.00 | 0.99 | 1.00 | 1.00 | 0.99 |
| 3 | 0.57 | 0.64 | 0.74 | 0.79 | 0.72 | 0.64 | 0.85 | 0.91 | 0.92 | 0.80 | 0.73 | 0.57 | 0.95 | 0.98 | 0.97 | 0.58 | 0.58 | 0.57 | 0.58 | 0.81 | 0.64 | 0.62 | 0.63 | 0.64 | 0.47 | 0.72 | 0.91 | 1.00 | | 1.00 | 1.00 | 1.00 | 1.00 | 1.00 | 1.00 | 1.00 | 1.00 | 1.00 | 0.99 |
| 17 | 0.55 | 0.63 | 0.73 | 0.78 | 0.71 | 0.63 | 0.84 | 0.91 | 0.91 | 0.79 | 0.72 | 0.56 | 0.94 | 0.97 | 0.96 | 0.56 | 0.56 | 0.56 | 0.56 | 0.80 | 0.62 | 0.60 | 0.61 | 0.62 | 0.46 | 0.71 | 0.90 | 1.00 | 1.00 | | 1.00 | 1.00 | 1.00 | 1.00 | 1.00 | 1.00 | 1.00 | 1.00 | 1.00 |
| 18 | 0.56 | 0.64 | 0.74 | 0.78 | 0.71 | 0.63 | 0.85 | 0.91 | 0.92 | 0.80 | 0.72 | 0.56 | 0.94 | 0.98 | 0.96 | 0.57 | 0.57 | 0.56 | 0.57 | 0.81 | 0.63 | 0.61 | 0.62 | 0.63 | 0.46 | 0.71 | 0.90 | 1.00 | 1.00 | 1.00 | | 1.00 | 1.00 | 1.00 | 1.00 | 1.00 | 1.00 | 1.00 | 1.00 |
| 25 | 0.55 | 0.63 | 0.73 | 0.77 | 0.70 | 0.62 | 0.84 | 0.90 | 0.91 | 0.79 | 0.71 | 0.55 | 0.94 | 0.97 | 0.96 | 0.56 | 0.56 | 0.55 | 0.56 | 0.80 | 0.62 | 0.60 | 0.61 | 0.62 | 0.45 | 0.70 | 0.90 | 1.00 | 1.00 | 1.00 | 1.00 | | 1.00 | 1.00 | 1.00 | 1.00 | 1.00 | 1.00 | 1.00 |
| 26 | 0.53 | 0.61 | 0.71 | 0.76 | 0.69 | 0.61 | 0.83 | 0.90 | 0.90 | 0.78 | 0.70 | 0.53 | 0.93 | 0.97 | 0.95 | 0.54 | 0.54 | 0.53 | 0.54 | 0.78 | 0.60 | 0.58 | 0.59 | 0.60 | 0.43 | 0.69 | 0.89 | 1.00 | 1.00 | 1.00 | 1.00 | 1.00 | | 1.00 | 1.00 | 1.00 | 1.00 | 1.00 | 1.00 |
| 27 | 0.54 | 0.62 | 0.72 | 0.77 | 0.70 | 0.62 | 0.84 | 0.90 | 0.91 | 0.78 | 0.71 | 0.54 | 0.93 | 0.97 | 0.96 | 0.55 | 0.55 | 0.54 | 0.55 | 0.79 | 0.61 | 0.59 | 0.60 | 0.61 | 0.44 | 0.70 | 0.89 | 1.00 | 1.00 | 1.00 | 1.00 | 1.00 | 1.00 | | 1.00 | 1.00 | 1.00 | 1.00 | 1.00 |
| 28 | 0.50 | 0.58 | 0.69 | 0.74 | 0.66 | 0.58 | 0.81 | 0.88 | 0.89 | 0.76 | 0.68 | 0.51 | 0.92 | 0.96 | 0.94 | 0.51 | 0.51 | 0.51 | 0.51 | 0.76 | 0.57 | 0.55 | 0.56 | 0.58 | 0.40 | 0.66 | 0.87 | 1.00 | 1.00 | 1.00 | 1.00 | 1.00 | 1.00 | 1.00 | | 1.00 | 1.00 | 1.00 | 1.00 |
| 29 | 0.50 | 0.58 | 0.68 | 0.73 | 0.66 | 0.57 | 0.81 | 0.88 | 0.88 | 0.75 | 0.67 | 0.50 | 0.91 | 0.96 | 0.94 | 0.50 | 0.51 | 0.50 | 0.51 | 0.76 | 0.57 | 0.55 | 0.56 | 0.57 | 0.40 | 0.66 | 0.86 | 0.99 | 1.00 | 1.00 | 1.00 | 1.00 | 1.00 | 1.00 | 1.00 | | 1.00 | 1.00 | 1.00 |
| 30 | 0.57 | 0.64 | 0.74 | 0.79 | 0.72 | 0.64 | 0.85 | 0.91 | 0.92 | 0.80 | 0.73 | 0.57 | 0.94 | 0.98 | 0.97 | 0.57 | 0.58 | 0.57 | 0.58 | 0.81 | 0.63 | 0.61 | 0.62 | 0.63 | 0.47 | 0.72 | 0.90 | 1.00 | 1.00 | 1.00 | 1.00 | 1.00 | 1.00 | 1.00 | 1.00 | 1.00 | | 1.00 | 0.99 |
| 37 | 0.50 | 0.58 | 0.69 | 0.74 | 0.66 | 0.58 | 0.81 | 0.88 | 0.89 | 0.75 | 0.67 | 0.50 | 0.92 | 0.96 | 0.94 | 0.51 | 0.51 | 0.50 | 0.51 | 0.76 | 0.57 | 0.55 | 0.56 | 0.57 | 0.40 | 0.66 | 0.87 | 1.00 | 1.00 | 1.00 | 1.00 | 1.00 | 1.00 | 1.00 | 1.00 | 1.00 | 1.00 | | 1.00 |
| 38 | 0.48 | 0.56 | 0.67 | 0.72 | 0.64 | 0.56 | 0.79 | 0.87 | 0.87 | 0.74 | 0.65 | 0.48 | 0.90 | 0.95 | 0.93 | 0.49 | 0.49 | 0.48 | 0.49 | 0.74 | 0.55 | 0.53 | 0.54 | 0.55 | 0.38 | 0.64 | 0.85 | 0.99 | 0.99 | 1.00 | 1.00 | 1.00 | 1.00 | 1.00 | 1.00 | 1.00 | 0.99 | 1.00 | |

Fig. 6. The correlation coefficient among voltage angle of buses in at t=4.15 sec

| | G1 | G2 | G3 | G4 | G5 | G6 | G7 | G8 | G9 | G10 |
|---|---|---|---|---|---|---|---|---|---|---|
| G1 | | 0.1 | 0.1 | 0.1 | 0.1 | 0.1 | 0.1 | 0.1 | 0.2 | 0.1 |
| G2 | 0.0 | | 0.2 | 0.1 | 0.1 | 0.1 | 0.1 | 0.1 | 0.1 | 0.1 |
| G3 | 0.0 | 0.2 | | 0.2 | 0.1 | 0.2 | 0.1 | 0.1 | 0.1 | 0.1 |
| G4 | 0.1 | 0.1 | 0.1 | | 0.2 | 0.1 | 0.1 | 0.0 | 0.0 | 0.0 |
| G5 | 0.1 | 0.1 | 0.1 | 0.2 | | 0.0 | 0.1 | 0.0 | 0.0 | 0.0 |
| G6 | 0.1 | 0.1 | 0.1 | 0.2 | 0.2 | | 0.3 | 0.0 | 0.0 | 0.0 |
| G7 | 0.1 | 0.1 | 0.1 | 0.1 | 0.1 | 0.2 | | 0.0 | 0.0 | 0.0 |
| G8 | 0.1 | 0.0 | 0.0 | 0.0 | 0.0 | 0.0 | 0.0 | | 0.1 | 0.2 |
| G9 | 0.2 | 0.0 | 0.0 | 0.0 | 0.0 | 0.0 | 0.0 | 0.2 | | 0.2 |
| G10 | 0.3 | 0.1 | 0.1 | 0.1 | 0.0 | 0.1 | 0.1 | 0.3 | 0.2 | |

Fig. 7. Ks Matrix at t=4.15 sec

Fig.6 demonstrates the correlation coefficient among voltage angle of buses at t=4.15 sec, showing that the proposed approach designs the cut in such a way that buses with higher coherency in frequency are located on the same island.

The procedure mentioned above determines the boundaries of the islands. According to Fig.5, the other cuts must be applied to the lines that carry power and are not belong to the island, i.e., lines 1-2, 4-5, 13-14 which has 1078 kVA, 496 kVA, and 713 kVA respectively. Cutting the edges between the three formed power islands will lead to a large load-generation imbalance in each island and a frequency drift in the islanded area. Case II is defined as an islanding solution for Case I, where the events are represented in Table 2. The last row in Table 2 is the only difference between the events in Case I and Case II, which are for partitioning the network based on the proposed approach.

Table 2. Events Occurred in Case II

| Time (sec) | Description |
|---|---|
| 2.00 | Short circuit on lines 3-4 |
| 2.40 | Switch event on lines 3-4 |
| 3.00 | Short circuit on lines 16-17 |
| 3.40 | Switch event on lines 16-17 |
| **4.15** | **Switch event on lines 1-2, 4-5, 13-14** |

The rotor angle of the system after separation (Case II) is shown in Fig.8 which indicates that the generators in all three islands are coherent and stable. Fig. 10 also shows that the islands are stable in terms of the frequency. Note that the over/under frequency relays are not modeled in this study and we do not consider the frequency bands. While the proposed cuts promise stable islands in terms of frequency, Fig.9. point to voltage instability on one island. This is due to the lack of reactive power sources on the island and can be resolved by taking the reactive compensation measures.

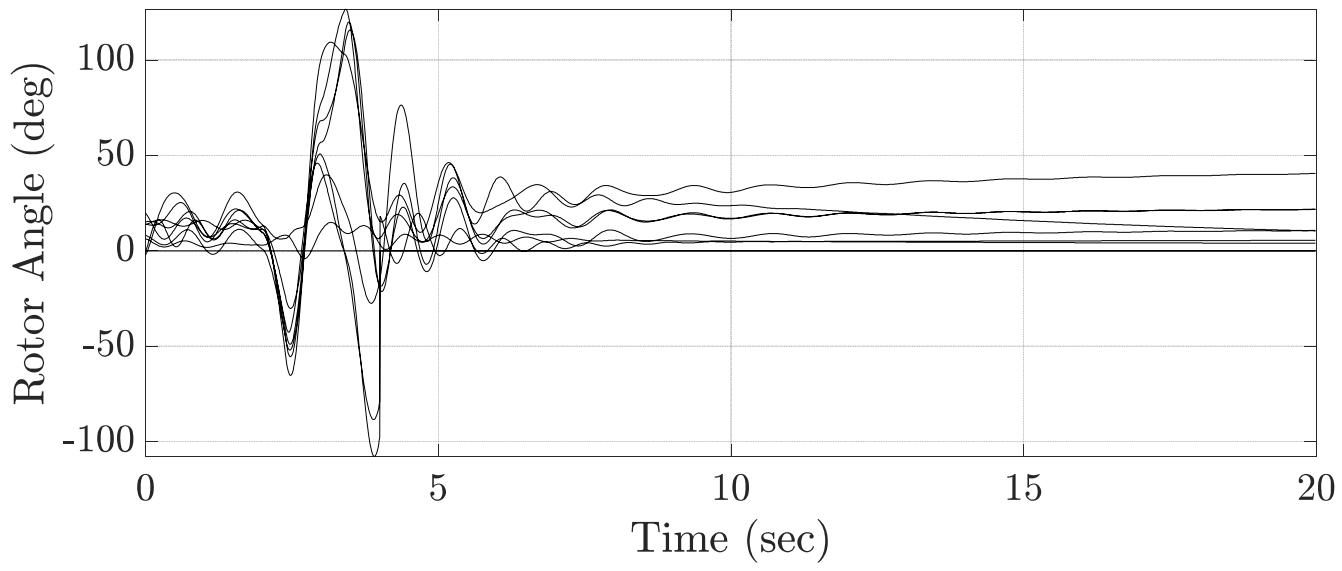

Fig. 8. Generator rotor angles in Case II

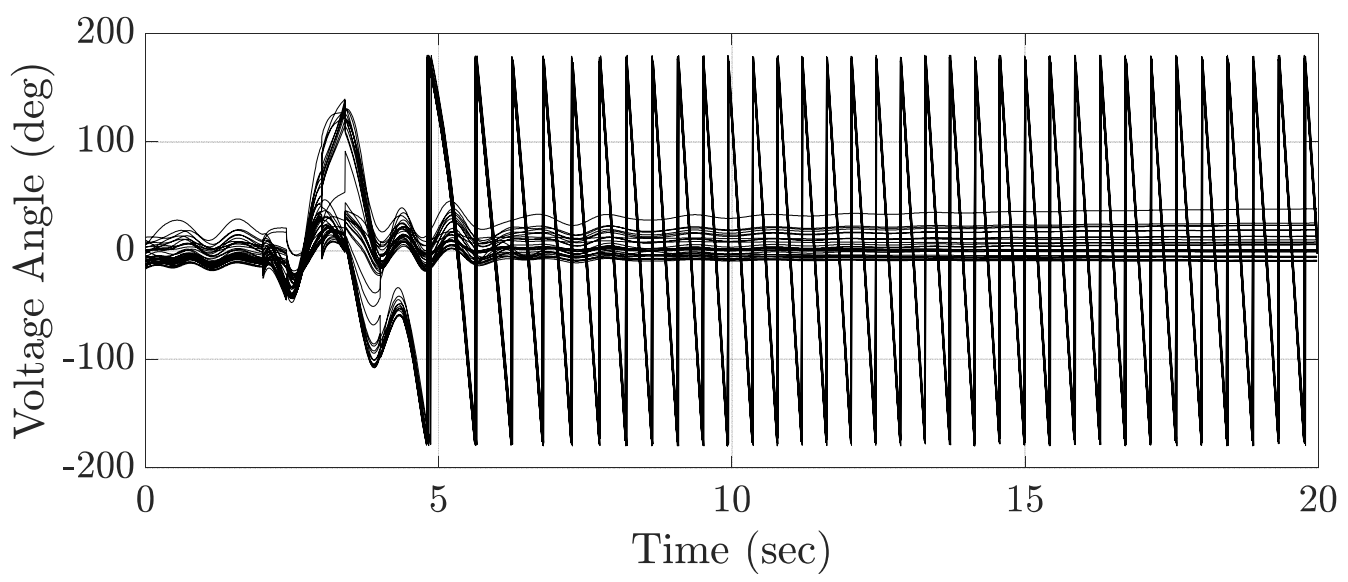

Fig. 9. Bus voltage angles in Case II

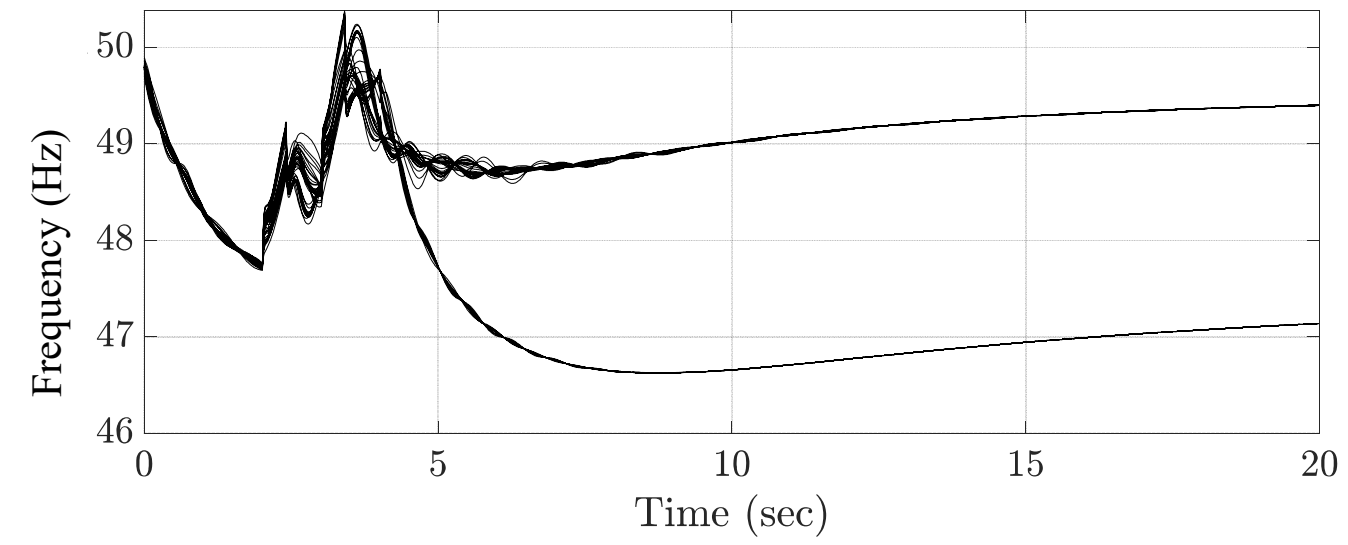

Fig. 10. Bus frequencies in Case II

## V. CONCLUSION

This paper proposes an efficient real-time intentional islanding algorithm based on the constraint clustering approach while addressing the generator coherency problem. We demonstrated that using the constraint clustering method, we can reach a stable solution when intentional islanding is needed. To this end, coherency among the generators is determine based on the synchronization coefficient, and the coherent generators are considered as the constraints for partitioning the buses. In other words, coherent groups of generators must be located on the same island. After obtaining a number of a coherent group of generators, the network is being partitioned based on the apparent power flow in the network branches. This problem is similar to find the minimum cut in the system which results as low power interruption in the line as possible.

It is shown that the use of the minimal power-flow disruption using apparent power flow improves the transient stability of the islands produced. The simulation results show that the proposed algorithm is computationally efficient and is suitable for use in real-time applications involving large power systems.

## REFERENCES

[1] Zhiqiang C., Zhihui D., Shuhuan W., Xuan L., Yanjun J., Linghao K., “The Application of Binary Particle Swarm Optimization in Power Restoration” presented Conference on Natural Computation 2014.

[2] P. A. Trodden, W. A. Bukhsh, A. Grothey, and K. I. M. McKinnon, “Optimization-based islanding of power networks using piecewise linear AC power flow,” IEEE Trans. Power Syst., vol. 29, no. 3, pp. 1212–1220, May 2014.

[3] K. Sun, D. Zheng and Q. Lu, "Splitting strategies for islanding operation of large-scale power systems using OBDD-based methods," IEEE Transactions on Power Systems, vol. 18, no. 2, pp. 912-922, May 2003.

[4] Ding T, Sun K, Huang C, Bie Z, Li F (2018) Mixedinteger linear programming-based splitting strategies for power system islanding operation considering network connectivity 12:350-359.

[5] Khalil AM, Iravani R (2016) A dynamic coherency identification method based on frequency deviation signals. IEEE Trans Power Syst 31:1779-1787.

[6] Ahad E., Mladen K., “Controlled Islanding to Prevent Cascade Outages Using Constrained Spectral k-Embedded Clustering” Power Systems Computation, 2016.

[7] Quirós-Tortós, J., Wall, P., Ding, L., and Terzija, V., 'Determination of Sectionalising Strategies for Parallel Power System Restoration: A Spectral Clustering-Based Methodology', Electric Power Systems Research, 2014, 116, pp. 381-390.

[8] L. Ding, F. M. Gonzalez-Longatt, P. Wall, and V. Terzija, “Two-step spectral clustering controlled islanding algorithm,” IEEE Trans. Power

[9] ZNIDI, F., DAVARIKIA, H., IQBAL, K. et al. Multi-layer spectral clustering approach to intentional islanding in bulk power systems. J. Mod. Power Syst. Clean Energy 7, 1044–1055 (2019) doi:10.1007/s40565-019-0554-1

[10] Sheikhrezaei, Kaveh. "Relating Individual Characteristics and Task Complexity to Performance Effectiveness in Individual and Collaborative Problem Solving." (2019).

[11] Sheikhrezaei, Kaveh, and Craig Harvey. "Relating the Learning Styles, Dependency, and Working Memory Capacity to Performance Effectiveness in Collaborative Problem Solving." In International Conference on Applied Human Factors and Ergonomics, pp. 53-64. Springer, Cham, 2019.

[12] Znidi, F., H. Davarikia, and K. Iqbal. Modularity clustering-based detection of coherent groups of generators with generator integrity indices. in Power & Energy Society General Meeting, 2017 IEEE. 2017. IEEE.

[13] Davarikia, H., Znidi, F., Aghamohammadi, M.R., Iqbal, K., “Identification of Coherent Groups of Generators Based on Synchronization Coefficient”, presented at the 2016 IEEE PES General Meeting, Boston, Massachusetts, July 17-21, 2016.

[14] H. Davarikia, M. Barati, F. Znidi, and K. Iqbal, "Real-Time Integrity Indices in Power Grid: A Synchronization Coefficient Based Clustering Approach", presented at the 2018 IEEE PES General Meeting, Portland, OR, July 17-21, 2018.

[15] P. A. Trodden, W. A. Bukhsh, A. Grothey, and K. I. M. McKinnon, “Optimization-based islanding of power networks using piecewise linear AC power flow,” IEEE Trans. Power Syst., vol. 29, no. 3, pp. 1212–1220, May 2014.

[16] L. Ding, F. M. Gonzalez-Longatt, P. Wall, and V. Terzija, “Two-step spectral clustering controlled islanding algorithm,” IEEE Trans. Power Syst., vol. 28, no. 1, pp. 75–84, Feb. 2013.

[17] Kamali S, Amraee T, Capitanescu F (2018) Controlled network splitting considering transient stability constraints 12:5639-5648

[18] Ding L, Ma Z, Wall P, Terzija V (2017) Graph spectra based controlled islanding for low inertia power systems. IEEE Trans Power Del 32:302-309.

[19] Khalil AM, Iravani R (2016) A dynamic coherency identification method based on frequency deviation signals. IEEE Trans Power Syst 31:1779-1787

[20] M.Jonsson, M. Begovic, and J. Daalder, "A new method suitable for real-time generator coherency determination," IEEE Transactions on the power system, vol. 19, pp. 1473-1482, 2004.